\documentclass[12pt,a4paper]{article}
\usepackage{eufrak,bbm,a4}
\newcommand{\frak}{\mathfrak}
\newcommand{\Bbb}{\mathbbm}

\begin{document}

\begin{center}

{\LARGE A certain class of \\

\vspace{0,1cm}

Einstein-Yang-Mills--systems \\}

\vspace{2cm}

{\Large Gerd Rudolph \footnote{e-mail: rudolph@rz.uni-leipzig.de} 
and Torsten Tok \footnote{e-mail: tok@tph100.physik.uni-leipzig.de}}

\vspace{0,3cm}

{\large Institute of Theoretical Physics, Leipzig University \\

\vspace{0,1cm}

04109 Leipzig, Augustusplatz 10}

\end{center}

\vspace{2cm}

\begin{abstract}

A class of $ G $-invariant Einstein-Yang-Mills (EYM) systems with 
cosmological constant on homogeneous spaces $ G / H $, where  $ G $ is a 
semisimple compact Lie group, is presented. These EYM--systems can be obtained 
in terms 
of dimensional reduction of pure gravity. If $ G / H $ is a symmetric space, 
the EYM--system on $ G / H $ provides a static solution of the EYM--equations 
on spacetime $ {\Bbb R} \times G / H $. This way, in particular, a 
solution for an arbitrary Lie group $ F $, considered as a symmetric space, 
is obtained. 
This solution is discussed in detail for the case $ F = SU(2) $.  
A known analytical EYM--system on $ {\Bbb R} \times S^3 $ is recovered 
and it is shown - using 
a relation to the BPST instanton - that this solution is of sphaleron 
type. Finally, a relation to the distance of Bures and to parallel 
transport along mixed states is shown.

%\noindent PACS: 04.20.Jb, 04.50.+h 

\end{abstract}

\newpage

\section{Introduction}
\label{intro}

In recent years there has been a lot of interest in static globally regular
finite energy solutions (so called soliton solutions) and
black hole solutions to Einsteins equations with Yang-Mills- , Higgs-
and other nonlinear sources. In 1988
Bartnik and McKinnon \cite{Bartnik} found a numerical solution
to the $ SU(2) $ Einstein-Yang-Mills (EYM) system. This was quite surprising,
because neither the vacuum-Einstein \cite{Lichnerowicz}
equations, nor the pure Yang-Mills 
\cite{Deser} equations on Minkowski space
have nontrivial soliton solutions.
Hence, the very weak gravitational interaction can change qualitatively the
spectrum of soliton solutions of a theory in which gravity was neglected.
Other authors discovered numerically black hole solutions
\cite{Bizon} to the $SU(2)$ EYM--system and the existence of both types of
solutions was established rigorously \cite{Smoller}. 
Unfortunately, all solutions to 
EYM--systems with arbitrary gauge group turned out to be unstable
in the sense of linear stability \cite{Straumann}. 
Subsequently several authors have investigated other models,
such as $SU(2)$ EYM-Higgs \cite{EYMH},
EYM-dilaton \cite{EYM-d} or Einstein-Skyrme systems \cite{ES} 
and found in some cases linearly stable solutions. 
Another line of research is to look for solutions of the EYM--equations 
in arbitrary spacetime dimension, 
which are invariant under some symmetry group K \cite{Rudolph2}. 
Then one has the possibility to apply the powerful theory of dimensional 
reduction and spontaneous compactification \cite{KMRV,Jadczyk}.

Some authors \cite{Hosotani,Ding,Molnar} discussed an
analytical $ SU(2) $ EYM--system in (3 + 1) dimensions with cosmological
constant $ \Lambda $. The aim of this paper is to show that 
this EYM--system is a special case of a
general construction of EYM--systems with arbitrary 
%semisimple compact 
gauge group.

The paper is organized as follows: In section \ref{general} we discuss 
a general 
construction, which yields on each homogeneous space $ G / H $ 
a $ G $-invariant EYM--system with cosmological 
constant. If $ G / H $ is a symmetric space, we can use this solution 
to obtain a static $G$-symmetric EYM--system on the spacetime 
$ {\Bbb R} \times G/H $. This is shown in section 
\ref{time-coupling}.
In section \ref{symmetric-space} we apply the construction to an arbitrary 
Lie group $ F $, considered as a symmetric space and in section 
\ref{symmetric-space-SU(2)} we discuss the special case $ F = SU(2) $. 
We recover the known analytical 
solution mentioned above \cite{Hosotani,Ding,Molnar}. Within our 
geometric framework we are able to give an intrinsic proof that the 
configuration under consideration is of sphaleron type. 
In the last section we show that this 
EYM--system appears also in the theory of parallel transport along mixed 
states.   

\section{General construction}
\label{general}

We consider a semisimple compact Lie group $ G $  and an arbitrary 
subgroup $ H 
\subset G $ with Lie algebras $ \frak G $ and $ \frak H $ resp.. 
On $ \frak G $ let their be given a positive definite $ Ad (G) $-invariant 
bilinear form $ \gamma $, for instance the negative of the Killing form $K$. 
The form $ \gamma $ defines by left transport a $G$-biinvariant 
metric $ \bf g $        
on $ G $ 
\begin{equation} 
\label{G-metric} 
{\bf g} ( {l_g}'X,{l_g}'Y ) = \gamma ( X , Y ) \quad , \quad
X,Y \in {\frak G} \quad .
\end{equation}
We denote by $l_g$ and $ r_g $ the left resp. right multiplication  
in the group $ G $ with group element $ g $. The prime denotes the
corresponding tangent map, i.e. $ {l_g}'X \in T_g G $. In terms of the 
canonical left invariant Lie-algebra-valued 
1-form $ \theta $ on $ G $ we can write symbolically 
\begin{equation}
\label{G-metric1}
\bf g = \gamma ( \theta  ,  \theta ) \, .
\end{equation}

Now we consider 
$ G $ as a principal bundle over $ G / H $ with structure group 
$ H $ and canonical projection $ \pi : G \rightarrow G / H$. 
Obviously $ \bf g $ is invariant under the right action of the 
structure group $ H $. Therefore, we can use the fact that  
every $ H $-invariant metric on an $ H $-principal bundle $ P $ 
with base space $ M $ 
defines and is defined by, three geometrical objects: 
A connection $ \Gamma $ in the bundle $ P $, a metric 
$ {\bf g}_{M} $ on the base space $ M $  
and, for every point $ x \in M $, an $ r_H $-invariant metric on  
$ E_x $, the fiber over $x$ \cite{Jadczyk}. 

We discuss these geometrical objects in our case. Let  $ P = G $, 
$ M = G / H $ and $ \bf g $ be the $ H $-invariant metric on $P$.  
We define for every $ g \in G $ the horizontal subspace $ Hor_g $ of the 
tangent space $ T_g G $ as the orthogonal complement  
 of the canonical vertical subspace $ Ver_g \subset T_g G $ of 
the bundle $ P = G $ with respect to the metric $ \bf g $. The vertical 
subspace $ Ver_g $ is obviously given by $ Ver_g = {l_g}' {\frak H} $. 
Thus, from (\ref{G-metric}) and the $ Ad (H) $-invariance of $ \bf g $ we see 
\begin{equation}
\label{hor}
Hor_g = {l_g}' \frak M \quad ,
\end{equation}
where $ {\frak M} \subset {\frak G} $ is the orthogonal 
complement of $ \frak H $ with respect to $ \gamma $. 
The decomposition 
\begin{equation} 
\label{reduct-decomp}
{\frak G} = {\frak H} \oplus {\frak M} 
\end{equation}
is because of the $ Ad (H) $ invariance of $ \gamma $ reductive, 
i.e. $[{\frak H} , {\frak M} ] \subset {\frak M } $.  
The horizontal subspaces $ Hor_g $ define a 
connection $ \Gamma $, which is known as the canonical connection 
in the bundle $ G(G / H , H) $ \cite{Kobayashi}. It's connection 
form $ \omega $ is given by the $ \frak H $-component of $ \theta $
with respect to the decomposition (\ref{reduct-decomp}):
\begin{equation}
\label{P-connection}
\omega = \theta_{\frak H} \, .
\end{equation}

The metric  ${\bf g}_{G/H} $ on the base space $ M = G / H $
can be obtained in the following way: Take two tangent vectors $ U_1 , U_2 
\in T_x (G / H) $ at the point $ x \in G / H $ and lift them to 
horizontal tangent vectors $ {\hat U}_1 , {\hat U}_2 \in Hor_g $ at an 
arbitrary point 
$ g \in E_x $, i.e. 
$ \pi' ( {\hat U}_i ) = U_i \, , i= 1,2 $. We put 
\begin{equation}
{\bf g}_{G/H} ( U_1 , U_2 ) := {\bf g} ( {\hat U}_1 , {\hat U}_2 ) \, .
\end{equation}
This definition 
of ${\bf g}_{G/H}$ is,
because of the $r_H$-invariance of $ \bf g $,
independent of the point $ g \in E_x $. 
If we have a (local) section $ s : G/H \rightarrow G $, 
then $ {\bf g}_{G/H} $ reads locally 
\begin{equation}
\label{M-metric}
{\bf g}_{G/H} = \gamma \left( (s^* \theta_{\frak M})  ,    
(s^* \theta_{\frak M}) \right) \, ,
\end{equation}
where $ \theta_{\frak M} $ is the $ {\frak M} $-component of $ \theta $ 
and the star denotes pull back. 
The $ r_H $-invariant metric on $ E_x $ is given by restriction of 
${\bf g}$ to $ E_x$. Due to the
$ l_G $-invariance of ${\bf g}$, every such a metric on $ E_x $ defines 
and is defined by 
the same $ Ad (H) $-invariant scalar product 
$ \gamma_{\frak H} $ on $ {\frak H} $, namely
\begin{equation}
\gamma_{\frak H} ( V_1 , V_2 ) = {\bf g} ( {l_g}' V_1 , {l_g}' V_2 ) \, ,
\end{equation}
where $ V_1 , V_2 \in {\frak H} $ and $ g \in G $ and  
 $ \gamma_{\frak H} $ is the restriction of 
$ \gamma $ to the subspace $ {\frak H} \subset {\frak G} $. 
One more consequence of the 
$ l_G $-invariance of $ \bf g $ is that the connection $ \Gamma $ and 
the metric 
${\bf g}_{G/H}$ are invariant under the left action of $ G $, compare with 
equations (\ref{P-connection}) and (\ref{M-metric}). 
 
Conversely, if we have a 
connection $ \Gamma $ with connection form $ \omega $ 
in the bundle $ G( G/H , H ) $ and a metric $ 
{\bf g}_{G/H} $ on $ G / H $, both not necessarily $l_G$-invariant,  and 
if we have an $ Ad (H) $-invariant  scalar product $ \gamma_{\frak H} $ 
on the Lie algebra $ \frak H $, then these three geometrical objects 
define a $ r_H $-invariant metric $ \bf g $ on $ G $ by \cite{Trautman}
\begin{equation}
\label{P-metric-constr}
{\bf g} ( X , Y ) := \gamma_{\frak H} ( \omega(X) , \omega(Y) ) + 
{\bf g}_{G/H} ( \pi'(X) , \pi'(Y) ) \, , X , Y \in T_g G \, .
\end{equation}

In the next step we will write down the scalar 
curvature $ R_G $ of the Levi-Civita connection on $ G $ in terms of 
$ \omega $, $ \gamma_{\frak H} $ and 
$ {\bf g}_{G/H} $. The $Ad (H)$-invariant scalar 
product $ \gamma_{\frak H} $ determines a biinvariant metric on $ H $, 
similarly as in equation (\ref{G-metric}). The scalar curvature of 
$ H $ calculated with respect to 
this metric is constant and is denoted by $ R_H $. 
The curvature $ \Omega $  
of the connection $ \Gamma $ is given by
\begin{equation}
\Omega = D \, \omega = d \, \omega + \omega \wedge \omega \quad .
\end{equation}
If we choose a local coordinate system $ ( x^\mu ) $ on $ G / H $ and 
a local section $ s : G/H \rightarrow G $ we can 
write 
\begin{equation} 
\label{F-mu-nu}
F = s^* \Omega = \frac{1}{2} F_{\mu \nu} \, d x^\mu \wedge d x^\nu \quad ,
\end{equation}
where $ F_{\mu \nu} $ takes values in $ \frak H $. 
Note that the  quantity $ \gamma_{\frak H} ( F_{\mu \nu} , F^{\mu \nu} ) $ 
considered as a function on $ G / H $ is   
independent of the section $ s $ and the coordinate system $ ( x^\mu ) $. 
Now the scalar curvature $ R_G $ reads, see  \cite{Jadczyk,KMRV}:
\begin{equation} 
\label{P-curvature}
R_G = \pi^* ( R_{G/H} ) + R_H - \frac{1}{4} \pi^* \left(
\gamma_{\frak H} ( F_{\mu \nu} , F^{\mu \nu} ) \right) \, ,
\end{equation}
where $ R_{G/H} $ denotes the scalar curvature on $ G / H $ and 
$ \pi^* $ the pull back under $ \pi $. 
Equation (\ref{P-curvature}) looks very simple,
but this is due to the $ Ad (H) $-invariance of $ \gamma_{\frak H} $ and 
the $ l_G $-invariant construction of 
the metric in the fibers $ E_x $, see equation (\ref{P-metric-constr}). 
In general the splitting of $ R_G $ is much more complicated.

It is clear that $R_G$ is constant on each fiber $ E_x $.
Therefore we can integrate the Einstein action
\begin{equation}
\label{Einstein-action} 
S = \int_G ( R_G - \Lambda_1 ) \, d v_G
\end{equation}
over the fibers $ E_x $ and we get
\begin{equation}
\label{EYM-action}
S = V_H \int_{G/H} \left( R_{G/H} - ( \Lambda_1 - R_H ) -
\frac{1}{4} \gamma_{\frak H} ( F_{\mu \nu} , F^{\mu \nu} )
\right) d v_{G/H} \quad . 
\end{equation}
Here $ d v_G $ resp.      
$ d v_{G/H} $  denote the volume forms on $ G $ resp. $ G / H $. 
$ V_H $ is the volume of the structure group $ H $, which is equal
to the volume of each fiber $ E_x $ and $ \Lambda_1 $ has the meaning 
of a cosmological constant. Equation (\ref{EYM-action}) 
gives the action of a coupled Einstein-Yang-Mills--system 
on $ G / H $ with cosmological constant $ \Lambda = \Lambda_1 - R_{H} $.

Hence, we arrive at the following result. If the metric $ \bf g $, given by 
(\ref{P-metric-constr}),   
is a solution of the Einstein equations 
with cosmological constant $ \Lambda_1 $, then the metric $ {\bf g}_{G/H} $ 
and the connection $ \Gamma $ form an Einstein-Yang-Mills--system 
with cosmological constant $ \Lambda = \Lambda_1 - R_H $.  
This is clear, because every variation of $ {\bf g}_{G/H} $ and 
$ \omega $ yields a variation of $ \bf g $. But $ \bf g $ is a solution 
of a variation principle with action (\ref{Einstein-action}). 
Therefore $ {\bf g}_{G/H} $ and $ \omega $ are solutions of 
a variation principle with action (\ref{EYM-action}).

It is a known fact that the biinvariant so called Killing metric 
$ {\bf g}_K $ on a  semisimple compact Lie group $ G $ 
arising from the negative  of the Killing form $ K $, 
see equation (\ref{G-metric}), 
is a solution of the Einstein 
equations with some cosmological constant. 
One can easily verify this statement by calculating the Ricci tensor $ Ric $. 
Doing this one gets \cite{Kobayashi}
%part II / chapter X / section 3 
\begin{equation} 
Ric =  \frac{1}{4} g_K \, . 
\end{equation}
Another consequence of this equation is that the scalar curvature $ R_G $ 
of $ G $ calculated with respect to $ {\bf g}_K $ is 
\begin{equation}
\label{G-curvature} 
R_G  = \frac{1}{4} D_G \, ,
\end{equation}
where $ D_G $ is the dimension of $ G $.
If we choose $ {\bf g} =  \alpha {\bf g}_K \, , \alpha > 0, $ as 
the metric on $ G $, the Einstein equations 
on $ G $ will be fulfilled. The corresponding cosmological constant 
is given by
\begin{equation}
\label{cosmological-constant}
\Lambda = \frac{D_G - 2}{4 \alpha} \, .
\end{equation}
Thus, we can use the above construction to find $G$-invariant EYM--systems 
on homogeneous spaces $ G / H $ with compact semisimple Lie group $ G $.

\section{EYM--systems on $ {\Bbb{ R}} \times G/H $}
\label{time-coupling}

Let $ G $ be a compact semisimple Lie group with Killing metric 
$ {\bf g}_K $ and $ H $ be a subgroup of $ G $,
so that $ G / H $ is a symmetric space. Then we can construct a
$ G $-invariant static EYM--system
on spacetime $ N = {\Bbb R} \times G / H $
using the construction described in the
foregoing section, with the first component of $ N $ playing the role 
of time. 

We use the Killing metric $ {\bf g}_K $ on $ G $ 
to obtain the 
EYM--system $ ( g_{G/H} , \omega_{G/H} ) $ on $ G / H $. 
Moreover, we have the projection $ \rho : N \rightarrow G / H $, 
which projects $ (t, x) \in {\Bbb R} \times G / H $ onto 
$ x \in G / H $. On $ N $ we consider the static metric 
\begin{equation}
\label{N-metric}
{\bf g}_N = - dt \stackrel{s}{\otimes} dt + \beta \, 
\rho^* ( {\bf g}_{G/H} ) \, , \,  \beta \in {\Bbb R}_+ \, , 
\end{equation}
and on the $ H $-bundle $ (\rho^* G) $ over $ G/H $ the static connection 
form  
\begin{equation}
\label{omega-N}
\omega_N = \rho^* ( \omega_{G/H} )        \, .
\end{equation}
Because the Yang-Mills equations are fulfilled on $ G/H $, one easily 
shows that the Yang-Mills equations on $ N $ are independently of 
$ \beta $ fulfilled, too. 
It remains to check the Einstein equations. 
If $ G / H $ is a {\it symmetric} space, it is easy to calculate 
the Ricci tensor $ Ric_{G/H} $ and the energy-momentum-tensor $ T_{G/H} $
on $ G / H $:  
\begin{eqnarray}
Ric_{G/H} &=& \frac{1}{2} {\bf g}_{G/H} \, , \\
T_{G/H} &=& \frac{4 - D_{G/H}}{8} {\bf g}_{G/H} \, ,
\end{eqnarray}
where $ D_{G/H} $ is the dimension of $ G / H $. Here we used 
\begin{equation}
{(T_{G/H})}_{\mu \nu} = \gamma_{\frak H} ( F_{\mu \tau} , {F_{\nu}}^\tau ) 
- \frac{1}{4} ({\bf g}_{G/H})_{\mu \nu} 
\gamma_{\frak H} ( F_{\tau \sigma} , F^{\tau \sigma} ) \ ,
\end{equation}
with the components $ F_{\mu \nu} $ of the field 
strength given by equation (\ref{F-mu-nu}). 

The Ricci tensor $ Ric_N $ on $ N $ calculated with respect to $ {\bf g}_N $ 
is, because of the simple structure of the metric $ {\bf g}_N $ 
(\ref{N-metric}),  given by 
\begin{equation}
\label{Ric-N}
Ric_N = \frac{1}{2} \rho^* ({\bf g}_{G/H}) , 
\end{equation}
i.e. $ Ric_N $ has no time components. The energy-momentum-tensor 
$ T_N $ on $N $ takes the form
\begin{equation}
\label{T-N}
T_N = \frac{1}{2 \beta} \rho^* ({\bf g}_{G/H}) - 
\frac{D_{G/H}}{8 \beta^2} {\bf g}_N \, .
\end{equation} 
Now it is a matter of fine tuning the constants to fulfill the Einstein 
equations on $ N $. The Einstein equations on $N$ read  
\begin{equation}
\label{Einstein-equation}
Ric_N - \frac{1}{2} {\bf g}_N \left( R_N - \Lambda_N \right) =
\kappa T_N \, ,
\end{equation}
where $ \Lambda_N $ is the cosmological constant and $ \kappa $ is 
the gravitational constant. Using equations (\ref{N-metric}), 
(\ref{Ric-N}) and (\ref{T-N})   
we calculate
\begin{eqnarray}
\nonumber Ric_N &-& \frac{1}{2} {\bf g}_N \left( R_N - \Lambda_N \right) = 
\frac{1}{2} \rho^* ({\bf g}_{G/H}) - \frac{1}{2} {\bf g}_N 
\left( \frac{D_{G/H}}{2 \beta} - \Lambda_N \right)  \\
\nonumber &=& \left( \beta \frac{\Lambda_N}{2} + \frac{2 - D_{G/H}}{4} \right) 
\rho^* ({\bf g}_{G/H}) +  dt \stackrel{s}{\otimes} dt 
\left( \frac{D_{G/H}}{4 \beta} - \frac{\Lambda_N}{2} \right) \\
\nonumber \kappa T_N &=& \left( \kappa \frac{4 - D_{G/H}}{8 \beta} \right) 
\rho^* ({\bf g}_{G/H}) + 
dt \stackrel{s}{\otimes} dt \left( \kappa 
\frac{D_{G/H}}{8 \beta^2} \right) 
\end{eqnarray}
Hence, we obtain two equations, which are equivalent to 
\begin{equation}
\label{fine-tuning-2}
\beta = \kappa \, , \, \Lambda = \frac{D_{G/H}}{4 \kappa} \, .
\end{equation}
Therefore, if the cosmological constant and the gravitational 
constant are related by equation (\ref{fine-tuning-2}), then 
the system $ ( {\bf g}_N , \omega_N ) $ is a solution of the EYM--equations 
on $N$.

This solution can also be obtained by solving the 
equations of spontaneous compactification \cite{KMRV} on 
$ {\Bbb R} \times G/H $.

\section{The symmetric space $(F \times F) / F_{diag}$}
\label{symmetric-space}   

Let $ F $ be a semisimple compact Lie group with Lie algebra $ \frak F $
and Killing form $ K_{\frak F} $. 
We can consider $ F $ as a symmetric space on which $ G = F \times F $ 
acts transitively by
$$ (f_1,f_2) * f := f_1 f {(f_2)}^{-1} \, , \, f_1,f_2,f \in F . $$
The stabilizer of the unit element $ e \in F $ is given by 
$$ F_{e} = \{ (f,f) ; f \in F \} = F_{diag} \equiv H \,  $$ 
and we can write
$$ F \equiv ( F \times F ) / F_{diag} = G / H \, . $$
The canonical splitting of the Lie algebra $ \frak G $ of $ G $ into a direct 
sum of the Lie algebra $ {\frak H} $ of $ H $ and a vector space 
$ \frak M $ is given by 
$$ {\frak G} \ni ( X , Y ) = \left( \frac{X + Y}{2} , \frac{X + Y}{2} \right) 
+ \left( \frac{X - Y}{2} , - \frac{ X - Y}{2} \right) \, , X,Y \in {\frak F}
\, ,  $$ 
i.e. 
\begin{eqnarray}
{\frak H} &=& \{ ( X , X ) ;  X \in {\frak F} \} \, , \\
{\frak M} &=& \{ ( X , - X ) ;  X \in {\frak F} \} \, .
\end{eqnarray}
The Lie algebra $ \frak G $ is the direct sum of two semisimple Lie 
algebras namely $ {\frak G} = {\frak F} \oplus {\frak F} $. Therefore, 
the Killing form $ K $ on $ \frak G $ is completely defined  
by $ K_{\frak F} $, namely
\begin{equation}
\label{Killing-split}
K \left( ( X_1 , X_2 ) , ( Y_1 , Y_2 ) \right) = 
K_{\frak F} \left(  X_1 , Y_1  \right) +
K_{\frak F} \left(  X_2 , Y_2  \right) \, .
\end{equation}
One easily obtains that  $ {\frak H}  $ and $ {\frak M} $ 
are orthogonal with respect to  $ K $.  

In the next step we  use $ \gamma = - \alpha K , \alpha > 0, $ 
to construct an EYM--system on $ F $ as described in section \ref{general}.
We denote  
the projection from $ G = F \times F $ onto the first resp. second 
component by $ pr_1 $ resp. $ pr_2 $ and we write 
$ \theta_i := pr_i^* \theta_F \, , i=1,2  $, where $ \theta_F $ is the 
canonical left invariant 1-form on $ F $. It is clear that $ G $ has the 
structure of a principal bundle over $ F $ with structure group 
$ H = F_{diag} $ 
and projection $ \pi : G \rightarrow F $ given by
\begin{equation}
\label{F-projection} 
\pi ( f_1 , f_2 ) = f_1 (f_2)^{-1} \, , f_1, f_2 \in F \, . 
\end{equation}
We choose the following 
section $ s : F \rightarrow G $ in this bundle 
\begin{equation}
\label{F-section}
s ( f ) := ( f , e ) \, , f \in F \, . 
\end{equation}
With these notations the canonical left 
invariant 1-form  $ \theta $ on $ G $ reads 
$$ \theta = (\theta_1 , \theta_2) \,  $$
and its projection onto $ {\frak H} $ resp. $ {\frak M} $ 
has the form 
\begin{eqnarray}
\label{theta-H-projection}
\theta_{{\frak H}} &=& \left( \frac{\theta_1 + \theta_2}{2} ,
\frac{\theta_1 + \theta_2}{2} \right) \, ,\\
\theta_{\frak M} &=& \left( \frac{\theta_1 - \theta_2}{2} , 
- \frac{\theta_1 - \theta_2}{2} \right) \, .
\end{eqnarray}
Moreover, we get 
\begin{eqnarray}
s^* \theta_{{\frak H}} &=& \left( \frac{\theta_F}{2} , 
\frac{\theta_F}{2} \right) \, ,\\
\label{s-stern-M}
s^* \theta_{\frak M} &=& \left( \frac{\theta_F}{2} ,
- \frac{\theta_F}{2} \right) \, ,
\end{eqnarray}
where we used $ pr_1 \circ s = id_F $ and $ pr_2 \circ s = e $. 
Now the gauge potential $ A_F = s^* \omega $
and the metric $ {\bf g}_F $, 
see equations (\ref{P-connection}) and (\ref{M-metric}), can be 
expressed by
\begin{eqnarray} 
\label{F-gaugefield}
A_F &=& s^* \omega = s^* \theta_{\frak H} = \left( \frac{\theta_F}{2} ,
\frac{\theta_F}{2} \right) \, , \\
\label{F-metric}
{\bf g}_F &=& \gamma \left( (s^* \theta_{\frak M})  ,
(s^* \theta_{\frak M}) \right) = 
- \frac{\alpha}{2} K_{\frak F} (\theta_F , \theta_F) \, . 
\end{eqnarray} 
To derive the last equation one has to take into consideration 
equations (\ref{s-stern-M}) and (\ref{Killing-split}). 
The scalar product $ \gamma_{{\frak H}} $ in the Lie algebra
$ {\frak H} $ of the structure group is given by the
restriction of $ \gamma $ to $ {\frak H} $. Identifying $ H = F_{diag} $ 
with $ F $ and using equation
(\ref{Killing-split}) we get
\begin{equation}
\label{structuregroup-scalar product}
\gamma_{{\frak H}} = - 2 \alpha K_{\frak F} \, . 
\end{equation}
Comparing with equation (\ref{G-curvature}), we obtain  
\begin{equation}
R_H = \frac{1}{8 \alpha} D_F \, ,
\end{equation}   
where $ R_H $ is the scalar curvature of the structure group and $ D_F $ 
is the dimension of $ F $. 
The metric $ \bf g $ on $ G = F \times F $ fulfilles the Einstein equations 
on $ G $ with cosmological constant $ \Lambda_1 = \frac{D_F - 1}{2 \alpha} $, 
see equation (\ref{cosmological-constant}). 
Thus, the cosmological constant $ \Lambda $ of the EYM--system 
consisting of $ {\bf g}_F $ and $ A_F $ reads
\begin{equation}
\label{cosm-const}
\Lambda = \Lambda_1 - R_H = \frac{3 D_F - 4}{8 \alpha} \, .
\end{equation}
The physical EYM-action has the form 
\begin{equation}
S = \int \left( \frac{1}{16 \pi \chi} (R - \Lambda) -
\frac{1}{4} ( F_{\mu \nu} , F^{\mu \nu} )    
\right) d v \, ,
\end{equation} 
where $ ( . , . ) $ is the negative of the Killing form and 
$ 8 \pi \chi = \kappa $ is the gravitational constant,
which appears in the Einstein equations (\ref{Einstein-equation}). 
It is obtained 
by dividing the action (\ref{EYM-action}) by $ 2 \alpha $ and identifying 
$ \alpha $ with the gravitational constant $ \kappa $.

Lets summarize our results. On every compact semisimple Lie group $ F $
exists an $ F $-symmetric EYM--system with gauge group $ F $
consisting of $ {\bf g}_F $ and $ A_F $, see equations (\ref{F-metric}) 
and (\ref{F-gaugefield}). The  gravitational constant  
$ \alpha $ and the cosmological constant $ \Lambda $  are 
related by equation (\ref{cosm-const}) (fine tuning).  

We considered $ F $ as a symmetric space. Therefore, we can apply 
the construction described in section \ref{time-coupling} to get 
an EYM--system on $ N = {\Bbb R} \times F $. 
We obtain the gauge potential and the metric on $ N $ 
from equations (\ref{omega-N}) and (\ref{N-metric}), 
where the EYM--system on $ F = G/H $ and 
the scalar product in the Lie algebra of the structure group $ F $
are given by equations 
(\ref{F-gaugefield}) and (\ref{F-metric}) 
resp. (\ref{structuregroup-scalar product}) with $ \alpha = 1 $. 
The 
resulting relations between the occuring constants $ \beta $ , 
$ \kappa $ and $ \Lambda_N $ can be obtained from equation 
(\ref{fine-tuning-2}).

\section{The case $ SU ( 2 ) \times SU ( 2 ) / SU( 2 ) $ - a relation to
instantons}
\label{symmetric-space-SU(2)}

For $ F = SU(2) $ it is easy to write down the explicit form
of the metric $ {\bf g}_{SU(2)} $ and the gauge potential $ A_{SU(2)} $ 
in local coordinates.  We  
parametrize $ SU(2) $ by stereographic coordinates $ z_\beta $
\begin{equation}
 SU(2) \ni {\bf x} = \frac{|z|^2 - 1}{|z|^2 + 1} {\Bbb 1} 
+ \frac{z_\beta}{|z|^2 + 1} i \sigma^\beta  ,
 z_\beta \in {\Bbb R} , \beta = 1,2,3  , 
|z|^2 = \sum_{\beta=1}^3 {(z_\beta)}^2  ,
\end{equation}
where $ \sigma^\beta $ denote the Pauli  matrices and $ \Bbb 1 $ is 
the $ 2 \times 2 $ unit matrix. 
The Killing form $ K_{su(2)} $ on the Lie algebra $ su(2) $ is given by
\begin{equation}
\label{su(2)-Killing-form}
K_{su(2)} ( X , Y ) =  4 tr ( X Y ) \, , X,Y \in su(2) \, . 
\end{equation}
To get the metric $ {\bf g}_{SU(2)} $ one has to use equation 
(\ref{F-metric})
\begin{equation}
{\bf g}_{SU(2)} = - \frac{\alpha}{2} K_{su(2)} ( \theta_{su(2)} , 
\theta_{su(2)} ) = - 2 \alpha tr \left( \theta_{su(2)} \stackrel{s}{\otimes} 
\theta_{su(2)} \right) \, .
\end{equation}
Here $ \stackrel{s}{\otimes} $ denotes the symmetrized tensor product and 
in what follows we write $ \bf \bar x $ for the hermitean conjugate of 
 $ \bf x $.
With $ \theta_{su(2)} = {\bf x}^{-1} d {\bf x} = - d ({\bf x}^{-1}) {\bf x} 
= - d {\bf \bar x} {\bf x} $ we get 
\begin{equation}
\label{Bures-metric}
{\bf g}_{SU(2)} = - 2 \alpha tr \left( d {\bf \bar x}  \stackrel{s}{\otimes} 
d {\bf x} \right) = \frac{16 \alpha}{(|z|^2 + 1)^2}  \sum_{\beta=1}^3
d z_\beta \stackrel{s}{\otimes} d z_\beta \, . 
\end{equation}
It is obvious that $ SU(2) $ endowed with this metric is a 3-sphere 
with radius $ 2 \sqrt{\alpha} $.
The gauge potential $ A_{su(2)} = \frac{1}{2} \theta_{su(2)} $ 
takes an especially simple form, if we 
perform a gauge transformation 
\begin{equation}
A' = u^{-1} A_{su(2)} u + u^{-1} d u 
\end{equation}
with 
\begin{equation}
u = \frac{{\bf \bar x} - {\Bbb 1}}{|{\bf x} - {\Bbb 1 }|} \, .
\end{equation}
After a simple calculation we get
\begin{equation}
\label{Uhlmann-connection}
A' = \frac{1}{|z|^2 + 1} z_\alpha d z_\beta
{\varepsilon^{\alpha \beta \gamma}} i \sigma_\gamma \quad ,
\end{equation}
where $ {\varepsilon^{\alpha \beta \gamma}} $ is totally antismmetric 
and $ {\varepsilon^{1 2 3}} = 1 $. The cosmological constant $ 
\Lambda $ follows from equation (\ref{cosm-const}) and has the value
\begin{equation}
\Lambda = \frac{3 D_F - 4}{8 \alpha} = \frac{5}{8 \alpha} \, .
\end{equation}
If we apply the construction described in section \ref{time-coupling}, 
then we obtain a static EYM--system on $ N = {\Bbb R} \times S^3 
\equiv {\Bbb R} \times SU(2) $.  
In local coordinates $ ( t , z_\alpha ) $ this solution reads
\begin{eqnarray}
A_N &=& \frac{1}{|z|^2 + 1} z_\alpha d z_\beta
{\varepsilon^{\alpha \beta \gamma}} i \sigma_\gamma \\
{\bf g}_N &=& - dt \stackrel{s}{\otimes} dt + 
\frac{16 \kappa}{(|z|^2 + 1)^2}  \sum_{\beta=1}^3
d z_\beta \stackrel{s}{\otimes} d z_\beta \, .
\end{eqnarray}
Here $ \kappa $ is the gravitational constant. 
From equations (\ref{structuregroup-scalar product}) and 
(\ref{su(2)-Killing-form}) we get 
the scalar product in the Lie algebra of the structure group  
\begin{equation}
\gamma_{su(2)} ( X , Y ) = - 8 tr ( X , Y ) \, , \, X , Y \in su(2) \, .
\end{equation}
The cosmological constant $ \Lambda_N $ we obtain from 
equation (\ref{fine-tuning-2}) and $ D_{SU(2)} = 3 $
\begin{equation}
\Lambda_N = \frac{3}{4 \kappa} \, .
\end{equation}
The same results were obtained in \cite{Ding,Molnar}. But in these 
papers the geometric structure of the solution was left in the dark.   
We hope that our considerations clarified this point completely. 

In paper \cite{Hosotani} there was mentioned a relation of the gauge 
potential $ A_{SU(2)} = \frac{1}{2} \theta_{SU(2)} $ to the 
BPST instanton solution \cite{BPST}, 
but only in terms of a local coordinate chart. 
In the bundle language, this relation looks as follows: 
Let us  consider 
the principal bundle $ P= SU(2) \times SU(2) \rightarrow SU(2) $ as 
a subbundle of the quaternionic Hopf bundle $ P_{\Bbb H } $: 
we show that the connection form $ \omega $ on $ P $, see equation 
(\ref{P-connection}),  is the pull back of the instanton connection 
form $ \omega_{inst} $ on $ P_{\Bbb H } $. 
The quaternionic Hopf bundle is given by
pairs $ ( a , b ) \, , \, a,b \in {\Bbb H } $ with
\begin{equation} 
\bar a a + \bar b b = 1 
\end{equation}
and by the right action of unimodular quaternions
\begin{equation} 
\psi_u ( a , b ) := ( a u , b u ) \, , \, \bar u u = 1 \, . 
\end{equation}
Here bar denotes quaternionic conjugation.
The set of unimodular quaternions is isomorphic to the group $ SU( 2 ) $.
Therefore, the bundle $ P = SU ( 2 ) \times SU ( 2 ) $ 
is naturally embedded by a bundle homomorphism $ i $
into the bundle $ P_{\Bbb H } $ 
\begin{equation}
i : P \ni ( a , b )  \rightarrow
( \frac{1}{\sqrt{2}} a , \frac{1}{\sqrt{2}} b ) \in P_{\Bbb H} \, .
\end{equation}
On $ P_{\Bbb H} $ the instanton connection is defined by
\begin{equation} 
\omega_{inst} = \bar a d a + \bar b d b \, . 
\end{equation}
Taking into account $ \bar u = u^{-1} $, for $u$ unimodular, we obtain 
\begin{equation}
i^* \omega_{inst} = \frac{1}{2} \left( \theta_1 + \theta_2 \right)  \, ,
\end{equation}
where $ \theta_i = {pr_i}^* \theta_{SU(2)} , i=1,2 $ as defined in the 
foregoing section. Comparing with equations (\ref{P-connection}) and   
(\ref{theta-H-projection}) it is clear that $ \omega $ is the pull 
back of $ \omega_{inst} $ under $ i $.
This gives us the possibility to calculate the Chern-Simons index $ k $ of 
the gauge potential $ A_{SU(2)} $, see equation (\ref{F-gaugefield}), 
in a very simple geometrically intrinsic way.  
If we represent the base space $ S^4 $ of the quaternionic Hopf bundle  
as the set of quaternions plus one point,  
we can choose the local section 
\begin{equation}
s_I : {\Bbb H} \ni x \rightarrow \left( \frac{x}{\sqrt{1 + |x|^2}} , 
\frac{1}{\sqrt{1 + |x|^2}} \right) \in P_{\Bbb H} 
\end{equation}
in the bundle $  P_{\Bbb H} $. We denote by $ A_I $ the 
instanton gauge potential and by $ F_I $ its field strength, 
i.e. $ A_I = {s_I}^* \omega_{inst} $.
Notice that the section $ s $ in the 
bundle $ P $, see equation (\ref{F-section}), is the restriction of 
$ s_I $ under the embedding $ i $.  Therefore, 
$ A_{SU(2)} $ is the pull back of $ A_I $ under $ i $. 
The embedding $ i $ induces an embedding of the base space of $ P $ 
into the base space of $ P_{\frak H} $, which we denote by the same 
letter $i $. 
The image $ i (M ) $ of the base space $ M = SU(2) $ of $ P $ 
is an equator of $ S^4 $. We denote one of the two halfspheres of $ S^4 $ 
whose boundary is $ i ( M ) $ by  $ N $. 
Now we can calculate 
\begin{eqnarray}
k & = & \frac{1}{8 \pi^2} \int_{SU(2)}
tr(A_{SU(2)} \wedge d A_{SU(2)} + \frac{2}{3} \,A_{SU(2)} \wedge 
A_{SU(2)} \wedge A_{SU(2)} ) \nonumber \\
& = & \frac{1}{8 \pi^2} \int_{\partial N}           
tr( A_I \wedge d  A_I + \frac{2}{3} \,A_I \wedge         
A_I \wedge A_I ) \nonumber \\
&=& \frac{1}{8 \pi^2} \int_{N} d \left( tr(A_{I} \wedge d A_{I} + \frac{2}{3}  
\,  A_{I} \wedge A_{I} \wedge A_{I} ) \right) \nonumber \\
&=& \frac{1}{8 \pi^2} \int_{N} tr ( F_{I} \wedge F_{I} ) \quad .
\end{eqnarray}
One easily checks that the connection 
$ A_I $ is up to gauge transformations 
invariant under the natural action of $ SO(5) $ on $ S^4 $.
 Therefore, $ tr ( F_I \wedge F_I ) $ is up to
a factor the volume form on $ S^4 $ and we get
\begin{eqnarray}
\nonumber
 \frac{1}{8 \pi^2} \int_{N} tr ( F_{I} \wedge F_{I} )
&=& \frac{1}{2} \frac{1}{8 \pi^2} \int_{S^4} tr ( F_{I} \wedge F_{I} )
\quad .
\end{eqnarray}
The topological index of the basic instanton is 1, hence the Chern-Simons
index $ k $ of our solution has to be $ 1/2 $. This is an  
intrinsic proof that the solution found is of sphaleron type.
The calculation of the Chern-Simons index in terms of local gauge 
potentials 
is much more complicated and may yield incorrect results, if one chooses 
a singular gauge. Therefore the authors in \cite{Ding} had to perform a 
gauge transformation before they had got the correct result.

\section[V]{A relation to Berry's phase and the distance of Bures}
\label{berry-bures}

The EYM--system $ ({\bf g}_{SU(2)} , A_{SU(2)} ) $, described in 
the foregoing section, is well known
from the study of parallel transport along $ 2 \times 2 $--
density matrices, see \cite{Rudolph1,Uhlmann,Tok}. 

We consider the trivial $ U(2) $ principal bundle
\begin{equation}
\label{U(2)-bundle}
GL(2,{\Bbb {C}}) \rightarrow GL(2,{\Bbb{C}}) / U(2) \equiv :  D_2(2)
\end{equation}  
with the projection $ \pi $ given by
\begin{equation}
\pi ( \omega ) := w w^\ast \, , w \in GL(2,{\Bbb {C}}) \,.
\end{equation}
Here star denotes hermitean conjugation. The base space $ D_2(2) $
consists of all not normalized nonsingular $ 2 \times 2 $ density matrices. 
On $ GL(2,{\Bbb {C}}) $ we have a natural metric 
$ \bf g $, which is invariant under right action of $ U(2) $
\begin{equation}
\label{GL(2)-metric}
{\bf g} = \Re \, tr \left( d \omega \stackrel{s}{\otimes} d \omega^* 
\right) \, .
\end{equation}
Therefore, we obtain by an analogous construction as in section 
\ref{general} a connection $ A_B $ in the bundle (\ref{U(2)-bundle}) and a 
metric $ {\bf g}_B $ on its base space.
The connection $ A_B $ was proposed by Uhlmann \cite{Uhlmann} and it 
governs parallel transport along mixed states, which is related to 
the concept of purification of density matrices. The metric $ {\bf g}_B $ 
is known as the Riemannian metric which comes from the distance of Bures 
\cite{Bures} and it is related 
to the transition probability between mixed states. 

It turns out that the connection $ A_B $ is reducible to a 
connection \cite{Rudolph1} in the $ SU(2) $ subbundle $  Q_2 $ 
defined by
\begin{equation}
Q_2 := \left\{ w \in GL(2,{\Bbb{C}}) ;
 det (w ) \in {\Bbb R}_+ \right\} \, .
\end{equation} 
This corresponds to a simple property of the metric (\ref{GL(2)-metric}):  
If we consider $ GL(2,{\Bbb C}) $ as a $ U(1) $ principal bundle 
over $ Q_2 $, then every vector tangent to $ Q_2 $ is orthogonal to the 
direction of the fiber of that bundle. Therefore, the horizontal 
subspaces of the connection $ A_B $, defined as the orthogonal complements 
of the vector spaces tangent to the fibers of the bundle 
$ GL(2,{\Bbb C}) \rightarrow D_2(2) $,  are tangent to $ Q_2 $. 
Another consequence of this 
property is that we can use the restriction 
of $ \bf g $ to $ Q_2 $ to construct $ {\bf g}_B $.

In the next step we restrict the base space of $ Q_2 $ to all 
normalized density matrices, 
\begin{equation}
tr ( \rho ) = tr ( w w^* ) = 1 \, .
\end{equation} 
We denote the resulting $ SU(2) $ bundle by $ \hat Q_2 $. Obviously, 
there is a natural embedding $ j : \hat Q_2 \rightarrow GL(2,{\Bbb C}) $.
On $ \hat Q_2 $ we have the pull back $ j^* {\bf g} $  
of the metric (\ref{GL(2)-metric}). 
We will show that there exists a bundle homomorphism 
$ f: \hat Q_2 \rightarrow P = SU(2) \times SU(2) $, such that 
$ j^* {\bf g} = f^* {\bf g}_P $, with $ {\bf g}_P $ being 
a multiple of the Killing metric on $ SU(2) \times SU(2) $. 

Obviously, every matrix $ w \in gl( 2 , {\Bbb C } ) $ can be
uniquely represented in the form 
\begin{equation}
\label{omega-P}
w = \frac{1}{2} \left( \begin{array}{cc}
1 & 0 \\
0 & i \\
\end{array} \right) \left( a + i b \right) \quad ,
\end{equation}
where 
\begin{eqnarray}
\nonumber
a &=& x_0 {\Bbb 1} + x_\alpha i \sigma^\alpha \quad , \\   
\nonumber
b &=& y_0 {\Bbb 1} + y_\alpha i \sigma^\alpha \quad ,  \alpha = 1,2,3 \quad ,
\quad x_i , y_i \in {\Bbb R} .
\end{eqnarray} 
If $ w \in \hat Q_2 $ we have $ det ( w ) > 0 $ and
$ tr ( w w^* ) = 1 $. Hence, we obtain 
\begin{eqnarray}
0 &=& \Im ( det ( w ) ) = \frac{1}{4} ( \sum_{i=0}^3 {x_i}^2
- \sum_{i=0}^3 {y_i}^2 ) \label{imag} \quad , \\
0 &<& \Re ( det ( w ) ) = \frac{1}{2} \sum_{i=0}^3 x_i y_i
\label{real} \quad , \\
1 &=& tr ( w w^* ) = \frac{1}{2} (\sum_{i=0}^3 {x_i}^2
+ \sum_{i=0}^3 {y_i}^2 ) \label{trace} \quad .
\end{eqnarray}
From equations (\ref{imag}) and (\ref{trace}) we find
\begin{equation}
\sum_{i=0}^3 {x_i}^2  =  \sum_{i=0}^3 {y_i}^2 = 1 
\end{equation}
and therefore $ a , b \in SU ( 2 ) $. So we can define a map
$ f: \hat Q_2 \rightarrow P $ according to equation (\ref{omega-P}).
One  easily checks that this map is an injective bundle homomorphism and     
because of
(\ref{real}) the set $ f(\hat Q_2 ) $ is an open subset of $ P $. 
Thus, for every $ ( a , b ) = p \in f ( \hat Q_2 ) $ with origin 
$ w = f^{-1} ( p ) $ the tangent space $T_p P $  
is isomorphic to the tangent space $ T_w \hat Q_2 $. 

With these remarks it is a matter  of a simple calculation to show 
$ j^* {\bf g} = f^* {\bf g}_P $. 
Let $ X_1 $ and $ X_2 $ be two vectors in $ T_w \hat Q_2 $ and 
$ ( A_1 , B_1 ) $ 
and $ ( A_2 , B_2 ) $ their images in  $ T_p P $, 
$p = (a,b) \in SU(2) \times SU(2)$. 
Taking into account equations (\ref{G-metric1}), (\ref{Killing-split}),  
(\ref{su(2)-Killing-form}) as well as  
$$ \theta (A_i,B_i) = (a^{-1}A_i , b^{-1} B_i ) \in su(2) \oplus su(2) 
\, , i=1,2 $$ 
and using 
$$ -(a^{-1} A ) = (a^{-1} A )^* = A^* a \,, \, a \in SU(2) \, ,$$  
we obtain
\begin{eqnarray}
{\bf g}_P \left( (A_1,B_1) , (A_2,B_2) \right) 
&=& - 4 \alpha \left( tr ( a^{-1} A_1  a^{-1} A_2 ) + 
tr (  b^{-1} B_1  b^{-1} B_2 ) \right) \nonumber \\ 
&=& 4 \alpha \left( tr ( A_1 {A_2}^* ) + tr ( B_1 {B_2}^* ) \right) \nonumber 
\\ &=& 4 \alpha \left( \Re \, tr ( (A_1 + i B_1) ({A_2}^* - i {B_2}^*) \right) 
\nonumber \\
&=& 16 \alpha \left( \Re \, tr ( X_1 {X_2}^* ) \right) \, ,
\end{eqnarray} 
showing $ j^* {\bf g} = f^* {\bf g}_P $, if $ \alpha = \frac{1}{16}$.

Hence, Uhlmanns connection reduced to a connection in $ \hat Q_2 $ and the 
pull back of the Bures 
metric to the space of nonsingular normalized density matrices coincide 
with the EYM--system presented in section \ref{symmetric-space-SU(2)}, 
see equations (\ref{Bures-metric}) and (\ref{Uhlmann-connection}).

\end{document}